\documentclass[twocolumn]{aa}			
\usepackage{graphicx}

\usepackage{txfonts}
\usepackage{natbib}
\bibpunct{(}{)}{;}{a}{}{,}
\newcommand{\msun}{$\rm M_{\odot}$}

\newcommand{\hi}{H\,{\sc i} }

\begin{document}
   \title{The molecular gas content of the
  advanced S+E merger NGC\,4441}

   \subtitle{Evidence for an extended decoupled nuclear disc?}

   \author{E. J\"utte
          \inst{1,2}
          \and
          S. Aalto\inst{3}
          \and
          S. H\"uttemeister
          \inst{1}
          }

   \offprints{eva.juette@astro.rub.de}
  \institute{
           Astronomisches Institut der Ruhr-Universit\"at Bochum,
            44780 Bochum, Germany\\
         \and
 Astron, 7990AA Dwingeloo, The Netherlands\\
             \and
             Onsala Space Observatory, Chalmers University of Technology, SE-439
92 Onsala, Sweden
             }

   \date{Received ; accepted }

 
  \abstract
   {Mergers between a spiral and an elliptical (S+E mergers) are poorly studied 
so far despite the importance for galaxy evolution. 
NGC\,4441 is a nearby candidate for an advanced remnant of such a merger, 
showing typical tidal structures like an
   optical tail and 
   two shells as well as two H\,{\sc i} tails. }
{The study of the  molecular gas content gives clues on
 the impact of the recent merger event on the star formation. Simulations
of S+E mergers predict contradictory scenarios concerning the strength and the extent
of an induced starburst.
 Thus, observations of the amount and the distribution
 of the
molecular gas, the raw material of star formation, are needed to understand
the influence of the merger on the star formation history. }
 {$^{12}$CO and $^{13}$CO (1-0) and (2-1)
observations were obtained using the Onsala Space Observatory 20\,m
and IRAM 30\,m telescope as well as the Plateau de Bure interferometer. 
These data allow us to carry out a basic analysis of the molecular
gas properties such as estimates of the molecular gas mass, its temperature and 
density and the star formation efficiency.}
{The CO observations reveal an extended
   molecular gas reservoir out to $\sim$ 4\,kpc, with a total
molecular gas mass of $\rm \sim 5\cdot 10^8\,M_{\odot}$. Furthermore, high resolution
imaging shows a central molecular gas feature, most likely a rotating disc
hosting most of the molecular gas ($\rm \sim 4\cdot 10^8\,M_{\odot}$). This nuclear disc shows a different sense of rotation than the large-scale \hi structure,  indicating a kinematically decoupled core.
We modeled
the state of the interstellar medium with the radiative transfer code
 {\tt RADEX}, using 
 the ratios of the $^{12}$CO and $^{13}$CO lines. The results
are consistent with a diffuse ($ n\leq 10^3\,$cm$^{-3}$) molecular
medium with no significant indications for cold and dense cores of ongoing star
formation. This is in agreement with the moderate star formation rate  of $\rm 1-2 M_{\odot}\ yr^{-1}$ found
in NGC\,4441. Under the assumption of a constant star formation rate, the gas depletion time is $\tau = 4.8 \cdot10^8$\,yr. NGC\,4441 might be a
nearby candidate for an early-type galaxy with a dominating A star population, a
 so-called E+A galaxy, being in a poststarburst phase
several $10^8$\,yr after a merger event. 
}
   {}

    \keywords{galaxies: interactions
-- galaxies: starburst
-- galaxies: individual: NGC\,4441
-- radio lines: galaxies
-- radio lines: ISM
               }
   \maketitle
%

\section{Introduction}
Following the $\Lambda$CDM models for galaxy evolution, mergers between
galaxies occur frequently and are essential to form the large galaxies 
we see today 
\citep[e.g.,][]{2003Ap&SS.284..325S}. 
While most observational studies concentrate on mergers between two spiral 
galaxies, simulations by \cite{2000ASPC..197..267N}, \cite{2001ASPC..230..453N}, 
\cite{2003ApJ...597L.117K} and \cite{2006MNRAS.372..839N} revealed the 
importance of mergers between a spiral and an elliptical (so-called S+E mergers)
 for the formation of bright ellipticals. Observations of this class of mergers
are needed to investigate their relevance and influence on galaxy evolution. 
\\
The well studied major mergers of two gas-rich disc galaxies can lead to 
ultraluminous
infrared galaxies (ULIRGs, $L_{FIR} > 10^{12}\,{\rm L_{\odot}}$,
\cite{1996ARA&A..34..749S}) because of 
a merger-induced extreme starburst. 
S+E mergers are
however poorly studied so far and thus, it is unclear   
if starbursts generally occur in these mergers and how they evolve. 
Besides the lack of observations,
models of S+E mergers are in strong disagreement concerning the prediction
of interaction-triggered enhanced star formation.
\cite{1993ApJ...405..142W} predict a congregation of the gas in
the center of the 
remnant galaxy, leading to a strong gas concentration and thus resulting in a
starburst, similar to ULIRGs but less  intense.
Simulations by \cite{1997ApJ...481..132K} however predict a dispersion of gas
clouds which might 
not lead to a starburst at all, because the density of the gas is too low for it
 to
collapse and form new stars. 
For the understanding of galaxy evolution it is necessary to know which 
scenario is more realistic and in particular how S+E mergers influence the
stellar population content in the remnant. Observations of the molecular
gas can show the amount and extent of raw material for star formation. Investigations of the molecular gas content in interacting galaxies showed a concentration of the gas towards the centre and an increase of the gas mass compared to non-interacting galaxies \citep[e.g.,][]{1992A&A...264..433B,1993A&A...269....7B, 1997A&AS..126....3H}. In ULIRGs, the molecular gas forms compact nuclear rotating rings and discs, fueling the central starburst \citep[e.g.,][]{1996ApJ...457..678B,1998ApJ...507..615D,1999AJ....117.2632B, 2006astro.ph.10378G}. It is not known so far whether this is also the case in S+E mergers.  
\\
One 'prototypical' S+E merger candidate is NGC\,4194, the Medusa. In the
optical we see a diffuse tail going to the north and on the opposite side two
stellar shells are visible, consistent with predictions. Molecular
gas is found  out to 4.7\,kpc away from the center,
i.e much more spread out than 
in the case of a ULIRG
\citep{2000A&A...362...42A,2001A&A...372L..29A}. However, this galaxy is
clearly undergoing an intense 
starburst phase, albeit not as intense as in ULIRGs
\citep{2004AJ....127.1360W}. 
\\
\noindent
Here we present CO maps of another
S+E merger candidate, NGC\,4441 (see Table\,\ref{n4441chap_intro}) \citep[see discussion of merger history in ][]{n4441hi}. 
The optical morphology of this galaxy is very similar to the Medusa. It possesses one
tidal tail and two bright shells on the opposite side which is typical for S+E merger remnants \citep[e.g.,][]{1984ApJ...279..596Q,1985LNP...232..151D, 1997ApJ...481..132K}.  The main body has an
elliptical shape \citep{1981A&A....97..302B} with a small dust lane through
the center along the minor axis. In contrast to the strong similarities in 
the optical morphology compared with the Medusa, the atomic gas distribution 
is significantly different, since the \hi\ forms two symmetric tidal tails 
 (compared to only one tail in the Medusa \citep{medusa}).
In NGC\,4441, the total \hi mass is $\rm 1.5\cdot 10^9\,L_{\odot}$ \citep{n4441hi}.
The ongoing star formation rate is rather low ($\rm 1-2\,M_{\odot}\ yr^{-1}$), because the merger is in such
an advanced phase that  most
of the gas has been already used for star formation  \citep{2005AIPC..783..343M,n4441hi}.
Using optical spectra \cite{1981A&A....97..302B} found indications for 
 a period of enhanced star
formation in the past, 
since the stellar population is younger than that of a normal
elliptical galaxy. Our own optical spectra confirm this and
we estimate that a moderate starburst occured $\sim$ 1\,Gyr ago 
\citep{2005AIPC..783..343M}. 
\\
\\
\begin{table}
      \caption[Basic properties of NGC\,4441]{Basic properties of
        NGC\,4441. The distance is based on
 ${\rm H_0 = 75\,km\ s^{-1}\ Mpc^{-1}}$.\\
$^b$ The Hubble type is taken from \cite{1995yCat.7155....0D}\\
$^b$ taken from \cite{n4441hi}.
        }
        \label{n4441chap_intro}
\centering
\begin{tabular}{lc}
\hline\hline
 property& \\
\hline
RA (2000) & 12:27:20.3\\
DEC (2000) & +64:48:05\\
$ v_{\rm opt,hel}$ (km\ s$^{-1}$) &  2722\\
$D$ (Mpc) & 36 \\  
type$^a$ & SAB0+ pec \\
$ L_{\rm B} \ (10^{9}\,{\rm L_{\odot}})$ &10.1 \\
$ L_{\rm FIR} \ (10^9\,{\rm L_{\odot}})$ & 5.4 \\
SFR$_{FIR}$ ($\rm M_{\odot}$) & 1.0 \\
SFR$_{\rm 20\,cm}^b$ ($\rm M_{\odot}$) & 2.4 \\
1\arcmin & 10.5\,kpc \\    
\hline
\end{tabular}
\end{table}

  \begin{figure}
   \centering
   \includegraphics[angle=0,width=8cm]{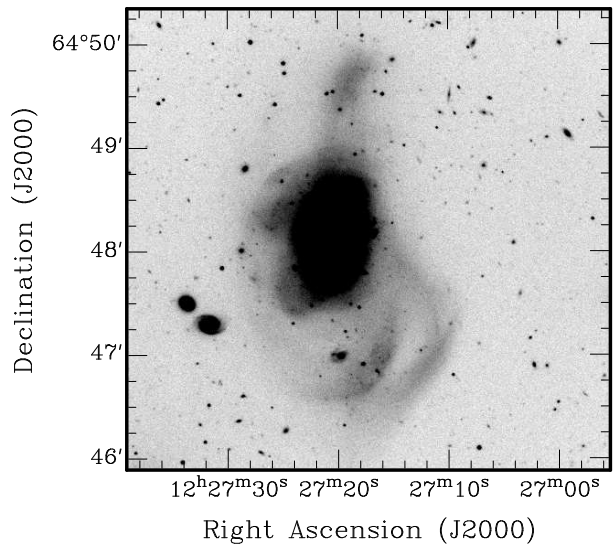}
  \caption{R-Band image of NGC\,4441, observed with the Calar Alto
2.2\,m, integration time 1\,h \citep{optsample}. Note the tidal tail to the 
north and two stellar shells to the south-west. The two galaxies south-east 
of the main body are background galaxies.  
}
  \label{n4441chap_opt}
   \end{figure}


\section{Observations \& data reduction}

\subsection{Onsala Space Observatory 20\,m}
First $\rm ^{12}$CO(1-0) observations were carried out in March 2003 using
the 20\,m 
telescope of the Onsala Space Observatory (OSO), Sweden. 
 CO 
observations were 
done with a SIS--mixer and a correlator with a total bandwidth of
512\,MHz. The main beam efficiency at 115\,GHz is $\rm \eta = 0.43$. To 
account for the highly variable sky background, the observations were done in
the beam switching mode.
 The
switching frequency was 1.8\,Hz and the beam throw 11\arcmin.
We observed the
center position as given in Table\,\ref{n4441chap_intro} under good weather
conditions, i.e. 30\% humidity and clear, stable sky. The pointing and focus was regularly checked during the
observations with the calibration sources R\,Leo and
R\,Cas. Table\,\ref{n4441chap_iramobs}  
summarises the observations. 

\subsection{IRAM 30\,m}
Follow-up CO observations of NGC\,4441 were done with the IRAM 
30\,m telescope on
Pico Veleta, Spain in July 2004. The weather conditions were good. 
We mapped an extended CO distribution both in the $\rm ^{12}$CO(1-0) and (2-1) 
line. 
We used the A\,100 and  B\,100 as well as the A\,230 and B\,230 receivers to
observe the $\rm ^{12}$CO(1-0) and (2-1) lines 
simultaneously. The 1\,MHz filterbank  with
$2\times512$ channels was chosen as a backend. The observations were done in 
beam 
switching mode. Pointing
and 
focus calibration were regularly checked by observing Saturn. As a second
pointing source near NGC\,4441, we observed the calibration source
0954+658. To get a proper spatial
coverage (a fully 
sampled map) even in the
$\rm ^{12}$CO(2-1) line, we mapped the center using a $3\times3$ grid with a
spacing of 
6\arcsec, 
 i.e. half the width of the $\rm ^{12}$CO(2-1) beam. For the outer positions, we 
used
steps of 22\arcsec (full $\rm ^{12}$CO(1-0) beamwidth) to cover a 
larger area, since we wanted to track the extent of the molecular gas.
In total, we observed 17 positions out to 44\arcsec. In the most distant
pointings we do not find emission any more, thus, we covered the
central molecular gas extent completely and are able to estimate source
size and total mass.\\
The central position was also observed in 
$\rm ^{13}$CO with the (1-0) and (2-1) transitions being measured simultaneously.

\begin{table}
\caption[IRAM 30\,m and OSO CO observations of NGC\,4441]{Parameters for the
  CO observations of NGC\,4441 with the OSO 20\,m and IRAM 30\,m telescopes
as well as the Plateau de Bure interferometer. The velocity resolution given 
is a smoothed one, which is used throughout this paper.}
\label{n4441chap_iramobs}
\centering
\begin{tabular}{lccc}
\hline\hline
 obs. parameters & IRAM 30\,m & OSO 20\,m & PdB\\
\hline
$T_{sys}$ (K)& 200-340 ($^{12}$CO) & $\sim$
 350 & 180\\ 
 & $\sim$ 145 ($^{13}$CO) & -- & --\\
$\theta_{beam}$ CO(1-0) &  22\arcsec& 33\arcsec & 3.25\arcsec\\
$\theta_{beam}$ CO(2-1) & 11\arcsec&--&--\\
$\Delta v$  CO(1-0) & 21.0\,km\ s$^{-1}$ & 33.8\,km\ s$^{-1}$
& 6.6\\ 
$\Delta v$  CO(2-1) & 21.0\,km\ s$^{-1}$ &-- & --\\  
$\eta$ &0.75 (1-0) / 0.52 (2-1) & 0.43& 0.6-0.9\\
\hline
\end{tabular}
\end{table}

\subsection{Plateau de Bure}
To map the CO distribution at higher spatial resolution we obtained  
interferometric observations in December 2005, May and July 2007 using 
the Plateau de Bure 
interferometer (PdBI). The observations were done with a dual polarization receiver covering 
a bandwidth of 4\,GHz with 240 channels and centered at the redshifted 
CO(1-0) line at 
114.252\,GHz. This leads to a resolution of 6.56\,km\ s$^{-1}$.
The beamsize was 3.25\arcsec $\times$ 2.65\arcsec, the position angle 122$^\circ$. The primary beam at 115GHz is 45\arcsec.
In total, we had an integration time of 16 hours on source The observations
were done with the 6Cq--E10, 5Dq--W05 and 5Dq configurations.
Flux, phase, and bandpass calibrators were
 1044+719, 0418+380,
3C\,273, and 3C\,454. 

\subsection{Data reduction}
The {\tt CLASS}\footnote{Continuum and Line Analysis Single-dish Software,
  http://www.iram.fr/IRAMFR/GILDAS/} package was used for the data reduction
at Onsala;  
the IRAM CO data were reduced using
{\tt XS}\footnote{ftp://yggdrasil.oso.chalmers.se/pub/xs/}, a graphical reduction
and analysis software for mm spectral line data written by P.
Bergman. After checking the quality of each single
spectrum, the data were averaged with a
  weighting based on
the 
system temperature and integration time. A first-order baseline was fitted to
the resulting spectrum and subtracted. The data were converted to main beam
brightness temperature ($\rm T_{MB}$)  
  using the beam efficiencies given in Table\,\ref{n4441chap_iramobs}. 
Finally, we smoothed
  the spectra  
to a velocity resolution of 21\,km\ s$^{-1}$ (IRAM) and 34\,km\ s$^{-1}$ (OSO),
 respectively,
to achieve a better signal-to-noise
  ratio in individual channels. For the different positions in the $\rm ^{12}CO(1-0)$ map we reached a noise
level of 1.6--5.3\,mK, in the $\rm ^{12}CO(2-1)$ map the noise level lies 
between 2.1 and 9.6\,mK $T_{\rm MB}$. For $\rm ^{13}CO$,
noise levels are 0.7\,mK (1-0) and 1.3\,mK (2-1).\\

The Plateau de Bure data were reduced using the {\tt GILDAS} task {\tt CLIC}.
After flagging and phase and flux calibration using the observed calibrator 
sources,
a {\tt uv}-table of the science data was created. The {\tt uv}-table was then 
transformed
into a map using a natural weighting scheme. This dirty map was CLEANed to correct 
for sidelobes. After CLEANing, the rms noise level in the cube
 is 2.9\,mJy\ beam$^{-1}$. Finally,
moment maps were created from the CLEANed cubes using a clip value of 
\linebreak
$4\,\sigma$ = 0.012\,Jy\ beam$^{-1}$ . 


\section{Results}

\subsection{CO distribution and molecular mass}
\begin{figure}
   \centering
   \includegraphics[angle=-90,width=10cm]{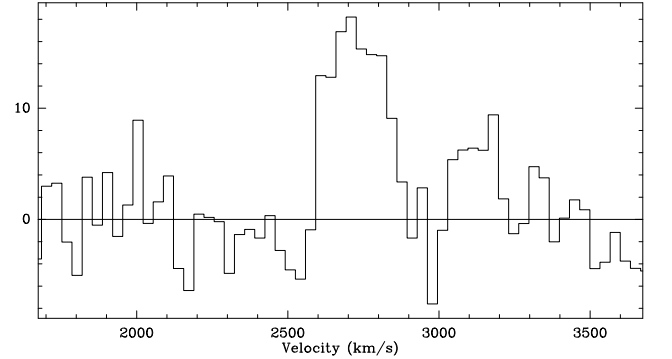}
  \caption[NGC\,4441: OSO $\rm ^{12}CO$ spectrum]{$\rm ^{12}$CO(1-0) spectrum
    observed with the OSO 20\,m 
    telescope. The intensity is given in $ T_{\rm MB}$ (mK). 
}
  \label{n4441chap_osospec}
   \end{figure}

First OSO observations revealed a molecular gas content in the inner 33\arcsec of
$\rm \sim 4.7\times 10^8\,M_{\odot}$.
The mass was derived by calculating the $\rm H_2$ column density
\begin{equation}
N({\rm H_2}) = X_{\rm CO} \cdot \int I_{\rm CO} {\rm d}v \ {\rm cm^{-2}}
\end{equation}
and thus
\begin{equation}
M_{\rm H2} = N({\rm H_2}) \cdot m({\rm H_2}) \cdot \Omega \ {\rm M_{\odot}}
\end{equation}
with $ X_{\rm CO}$ the CO-${\rm H_2}$ conversion factor, 
$I_{\rm CO}$ the intensity of the CO line in Kelvin,
 $m({\rm H_2})$ the mass of an $\rm H_2$ molecule in kg and $\Omega$ the
area covered by the beam in linear scales (cm$^2$).
 We used a 'standard' 
conversion
factor of 
$ X_{\rm CO} = 2.3 \cdot 10^{20}{\rm \,cm^{-2}(K\,km\ s^{-1})^{-1}}$,
referring to \cite{1988A&A...207....1S}.
Fig.\,\ref{n4441chap_osospec} shows the observed CO spectrum.  \\

\begin{figure*}[h!]
\centering
   \includegraphics[angle=270,width=15cm]{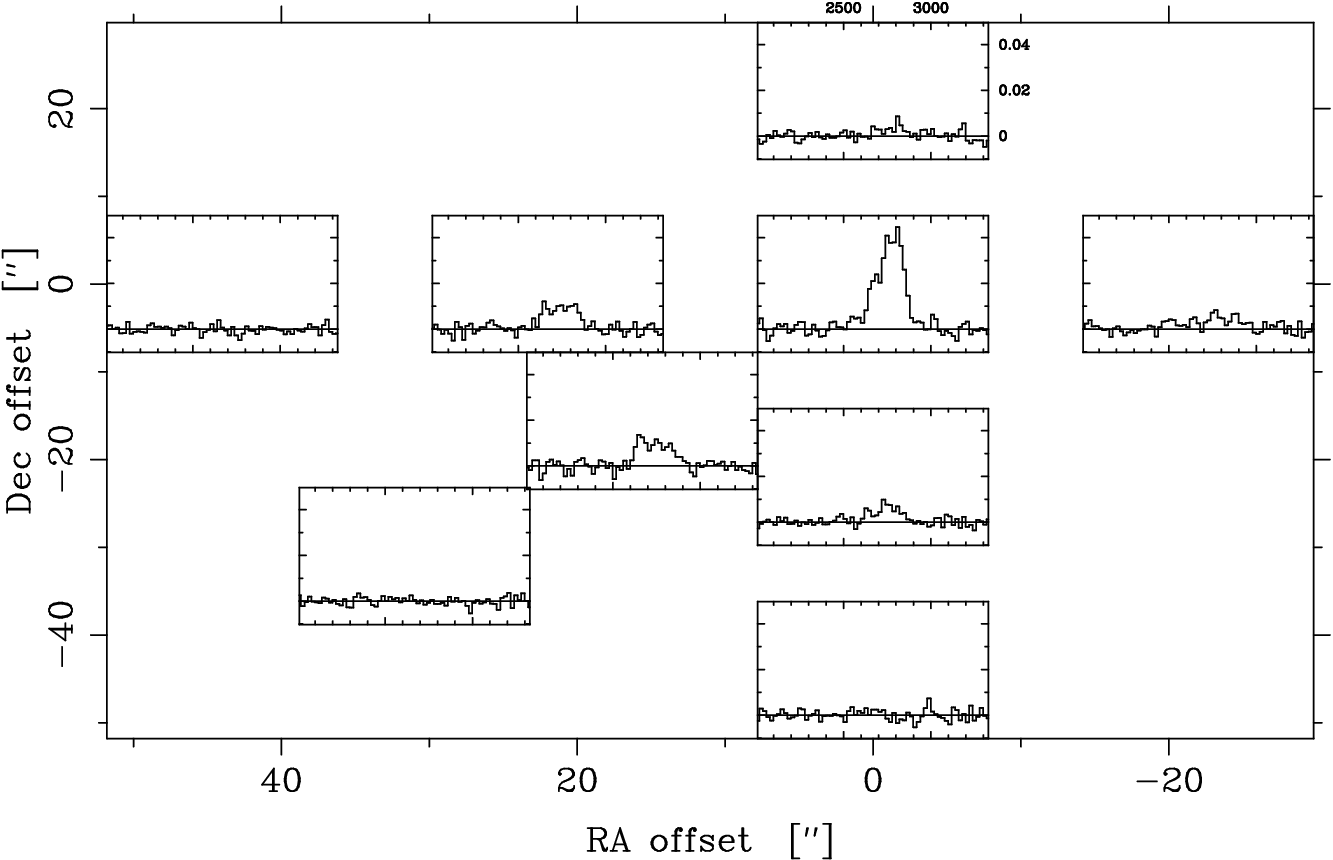}
    \caption[NGC\,4441: Large $\rm ^{12}$CO(1-0) map ]{Large $\rm
              ^{12}$CO(1-0) map of NGC\,4441, observed with the IRAM 
              30\,m 
              telescope. The outer tick marks denote the spacing of the observed
              positions, the inner tick marks represent the velocity (km\ s$^{-1}$) and $
              T_{\rm MB}$ (mK), respectively. 
              }
         \label{n4441chap_co10map}
   \end{figure*}

\begin{figure}[h!]
   \centering
   \includegraphics[angle=270,width=8cm]{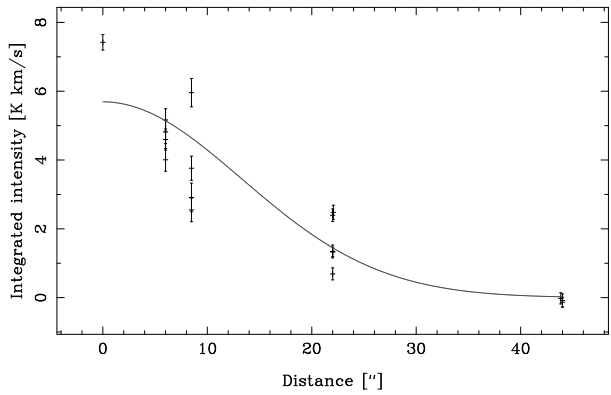}
      \caption{The observed integrated intensities of $^{12}$CO(1--0) 
are plotted versus the 
distance from the central position. To estimate the molecular source size 
we fitted a Gaussian and determined the FWHM = 22 $\pm$ 1\arcsec.
              }
         \label{scatter}
   \end{figure}

Mapping NGC\,4441 in CO with the IRAM 30\,m telescope, we found a molecular
gas distribution  extended in particular to the 
 south-east. Table\,\ref{n4441chap_irampoint} gives the pointings relative to
 the center 
position, the noise levels, the widths determined from Gaussians of both the $\rm ^{12}$CO(1-0) and CO(2-1) 
lines and the integrated 
intensities. In Fig.\,\ref{n4441chap_co10map} and
Fig.\,\ref{n4441chap_co21map} we 
present the spectra of the mapped positions for both the $\rm ^{12}$CO(1-0) and
CO(2-1) lines.
\\
Under the assumption of a Gaussian distribution of the intensity, we plotted
the intensity
versus radial distance from  
the center and 
fitted a Gaussian to derive the source size (see Fig.\,\ref{scatter} for
 $\rm ^{12}$CO(1-0)). 
The real source size can be
estimated from the observed source size and the known beam size as 
${\rm \Theta^2_{source} = \Theta^2_{obs} - \Theta^2_{beam}}$.
We derive a deconvolved source size of FWHM\footnote{full width of half maximum}=22\arcsec\ $\pm$ 1\arcsec (3.9\,kpc) for $\rm ^{12}$CO(1-0).
 The estimated
source 
size for $\rm ^{12}$CO(2-1) is slightly larger (30\arcsec $\pm$ 3\arcsec), but this is
probably an 
artefact due to undersampling of the source coverage and lower
signal-to-noise ratios in this
transition.  In our further discussion,
 we assume an
identical extent of  $\rm ^{12}$CO(2-1) and (1-0).
 We determined a central $\rm H_2$ column density of $\rm 1.8\cdot
 10^{21}\,cm^{-2}$ 
and calculated a total molecular gas
mass 
of $\rm 4.6\cdot 10^8\,M_{\odot}$, according to equations (1) and (2). \\
Since we found extended gas to the south and east (see
Fig.\,\ref{n4441chap_co10map} and Fig.\,\ref{n4441chap_co21map}) but only 
tentative detections to the north and west, the source seems to be asymmetric.
However, the estimated mass is in good agreement with the total mass derived from the OSO observations, which due to the larger beam cover the whole area of CO emission.

\begin{table*}[h!]
\caption[NGC\,4441: CO properties of the IRAM 30\,m map]{CO line parameters of
  the map of 
  NGC\,4441. The reference 
  coordinates are the center coordinates (J2000): RA: $\rm 12^h27^m20.36^s$,
  DEC: 
  $\rm +64^{\circ}48'06''$. Given are the offset positions, the total
  velocity range of the 
  CO(1-0) 
  and CO(2-1) 
  line and the intensities of both lines. The temperatures are measured in
  $T_{\rm MB}$.} 
\label{n4441chap_irampoint}
\centering
\begin{tabular}{lccccccc}
\hline\hline
$\Delta$ RA & $\Delta$ DEC & $\rm rms_{CO(1-0)}$ & ${\rm \Delta v_{CO(1-0)}}$
& 
 ${\rm I_{CO(1-0)}}$ & $\rm rms_{CO(2-1)}$
  & ${\rm \Delta v_{CO(2-1)}}$  & ${\rm I_{CO(2-1)}}$ \\
 ($''$)& ($''$)&(mK)&(km\ s$^{-1}$)& (K\,km\ s$^{-1}$)& (mK)  &(km\ s$^{-1}$) &(K\,km\ s$^{-1}$)  \\
\hline
0 &0 & 2.50& 292& 7.52& 7.67&324& 6.10 \\
6 &6 & 4.62& 283& 2.34& 6.72&240& 1.84 \\
0 &6 & 3.43& 328& 4.80& 6.34&270& 3.51 \\
-6 &6 & 5.33& 209& 2.68& 5.95&252& 2.69 \\
6 &0 & 3.78& 267& 4.24& 7.67&302& 3.64 \\
-6 &0 & 3.86& 336& 1.69& 7.67&247& 2.50 \\
6 &-6 & 5.28& 344& 6.65& 7.87&220& 6.34 \\
0 & -6 & 3.75& 320& 4.60& 9.6&285& 6.14  \\
-6 & -6& 4.42& 259& 3.70& 9.6&308& 3.97  \\
0 &22 & 2.19& 220& 0.92& 3.62&--& -- \\
-22 &0 & 2.06& --& 1.67& 2.88&--& -- \\
0 &-22 & 1.64& 262& 1.44& 3.07&--& -- \\
22 &0 & 2.33& 316& 2.47& 2.30&275& 0.88 \\
15.6 &-15.6 & 2.39& 310& 2.53&2.11 &318& 2.23 \\
44 & 0 & 1.77& --& --& 3.07&--& -- \\
0 & -44 & 2.18&-- & --& 2.11&--& -- \\
31 & -31 & 2.00& --& --& 2.88&--& -- \\
\hline
\end{tabular}
\end{table*}

\begin{figure*}[h!]
   \centering
   \includegraphics[angle=270,width=15cm]{co21_large.ps}
      \caption[NGC\,4441: Large $\rm ^{12}$CO(2-1) map ]{Large $\rm
              ^{12}$CO(2-1) map of NGC\,4441, observed with the 
              IRAM 
              30\,m 
              telescope. The outer tick marks denote the spacing of the observed
              positions, the inner tick marks represent the velocity (km\ s$6{-1}$) and $
              T_{\rm MB}$ (mK), respectively. 
              }
         \label{n4441chap_co21map}
   \end{figure*}

\begin{figure*}[h!]
   \centering
   \includegraphics[angle=-90,width=15cm]{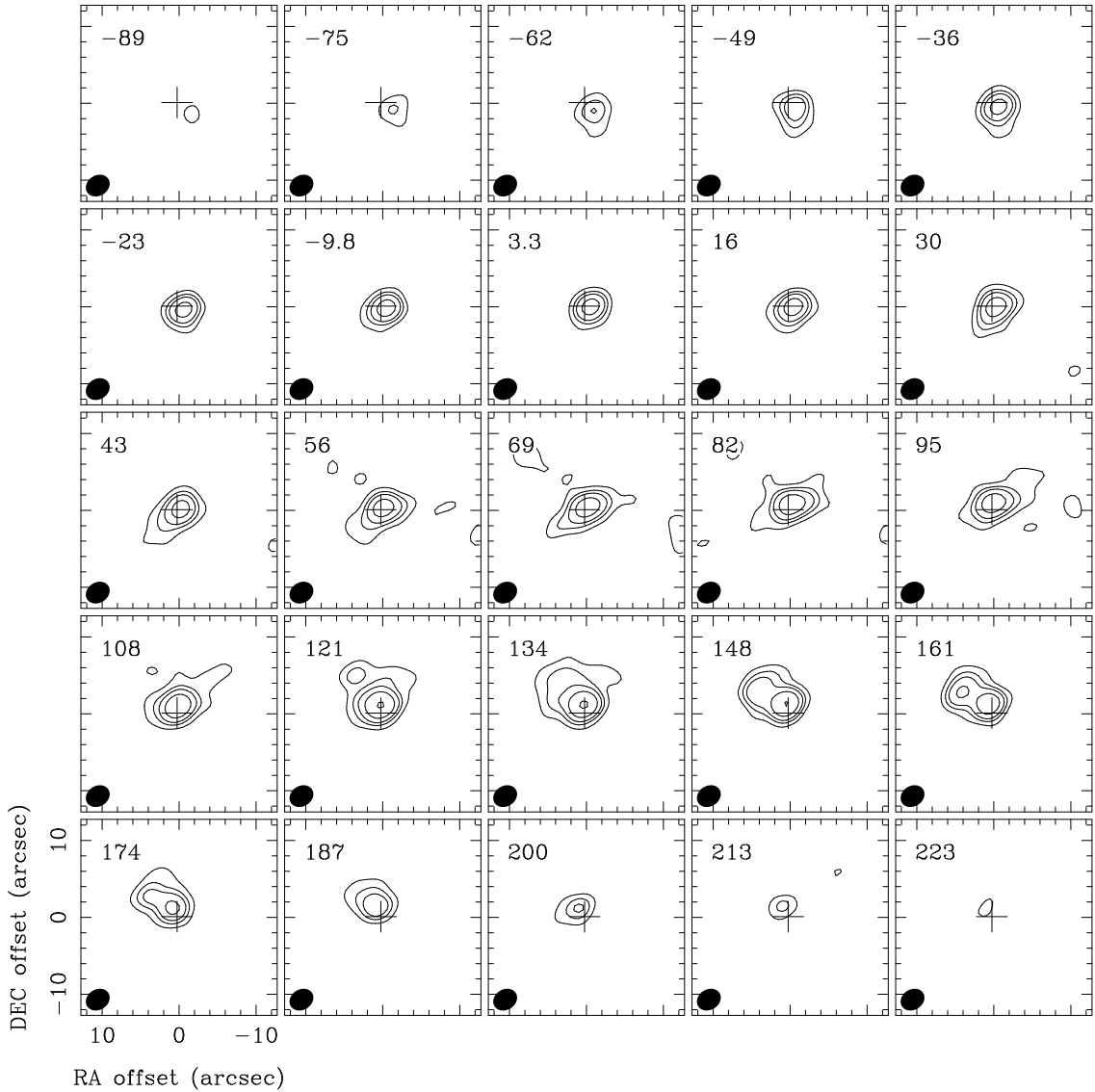}
      \caption{Channel maps of the nuclear rotating molecular gas disc observed
with the Plateau de Bure interferometer. The contours are $\rm 1,2,3,5,10,15 \cdot
0.01 Jy\,beam^{-1}$. The central velocity is 2652\,km\,s$^{-1}$.
              }
         \label{co_chanmap}
   \end{figure*}

\subsection{Molecular gas kinematics}

\begin{figure}[h!]
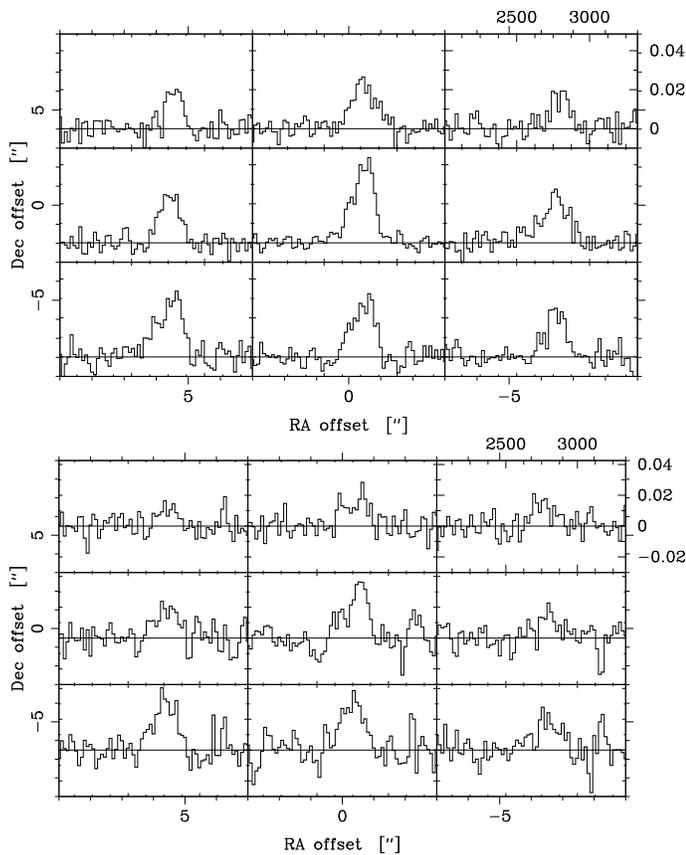

   \centering
   \includegraphics[angle=270,width=9cm]{co1-0_center_neu.ps}
   \includegraphics[angle=270,width=9cm]{co2-1_center_neu.ps}
   \caption[NGC\,4441: Inner 6\arcsec $\times$ 6\arcsec maps in $\rm ^{12}$CO(1-0)
   and $\rm ^{12}$CO(2-1) ]{Inner 6\arcsec $\times$ 6\arcsec of the CO map observed
   with the IRAM 
   30\,m telescope. The outer tick marks denote the spacing of the observed
              positions, the inner tick marks represent the velocity (km\ s$^{-1}$) and $
              T_{\rm MB}$ (mK), respectively (labeled in the upper right
              spectrum). {\bf top:} $\rm ^{12}$CO(1-0) map of the central
   region. Note that the data are highly oversampled (beamsize 22\arcsec). {\bf
   bottom:} $\rm ^{12}$CO(2-1) map of the central region. Note that the data
   are close to being fully sampled (beamsize 11\arcsec).  See Fig\,
   \ref{n4441chap_co10map} and 
   \ref{n4441chap_co21map} for the larger scale maps. }
   \label{n4441chap_cocenter}
\end{figure}

The spectra in both maps (Fig.\,\ref{n4441chap_cocenter},
\ref{n4441chap_co10map}, \ref{n4441chap_co21map}) clearly show  two
components, a fainter one (measured at the 
center) 
at $v_{\rm lsr}=2690 \pm$ 15\,km\ s$^{-1}$ (2660 $\pm$ 18\,km\ s$^{-1}$ in $\rm ^{12}$CO(2-1)) and the dominant one at
$v_{\rm lsr}= 2790 \pm$ 10\,km\ s$^{-1}$.  
Comparing the spectra at individual positions, the relative intensities of
both line components change, in particular in the SE direction. In
the 
$\rm ^{12}$CO(2-1) map, the more blueshifted line is the brighter one at the 
offset
position (+15.6\arcsec, -15.6\arcsec) (see Fig.\,\ref{n4441chap_co21map}).
 This component is centered at 2660\,km\ 
s$^{-1}$ in
$\rm ^{12}$CO(2-1) in each position, whereas in $\rm ^{12}$CO(1-0) there seems 
to be a
slight 
shift from 2690\,km\ s$^{-1}$ in the center to 2660\,km\ s$^{-1}$ in the outer
region. \\
Generally, the single dish molecular gas kinematics agrees with the overall 
central
kinematics of the H\,{\sc i} presented in \cite{n4441hi} in terms of the 
covered velocity range.

\subsection{The nuclear region}

\begin{figure}[h!]
   \centering
   \includegraphics[angle=0,width=8cm]{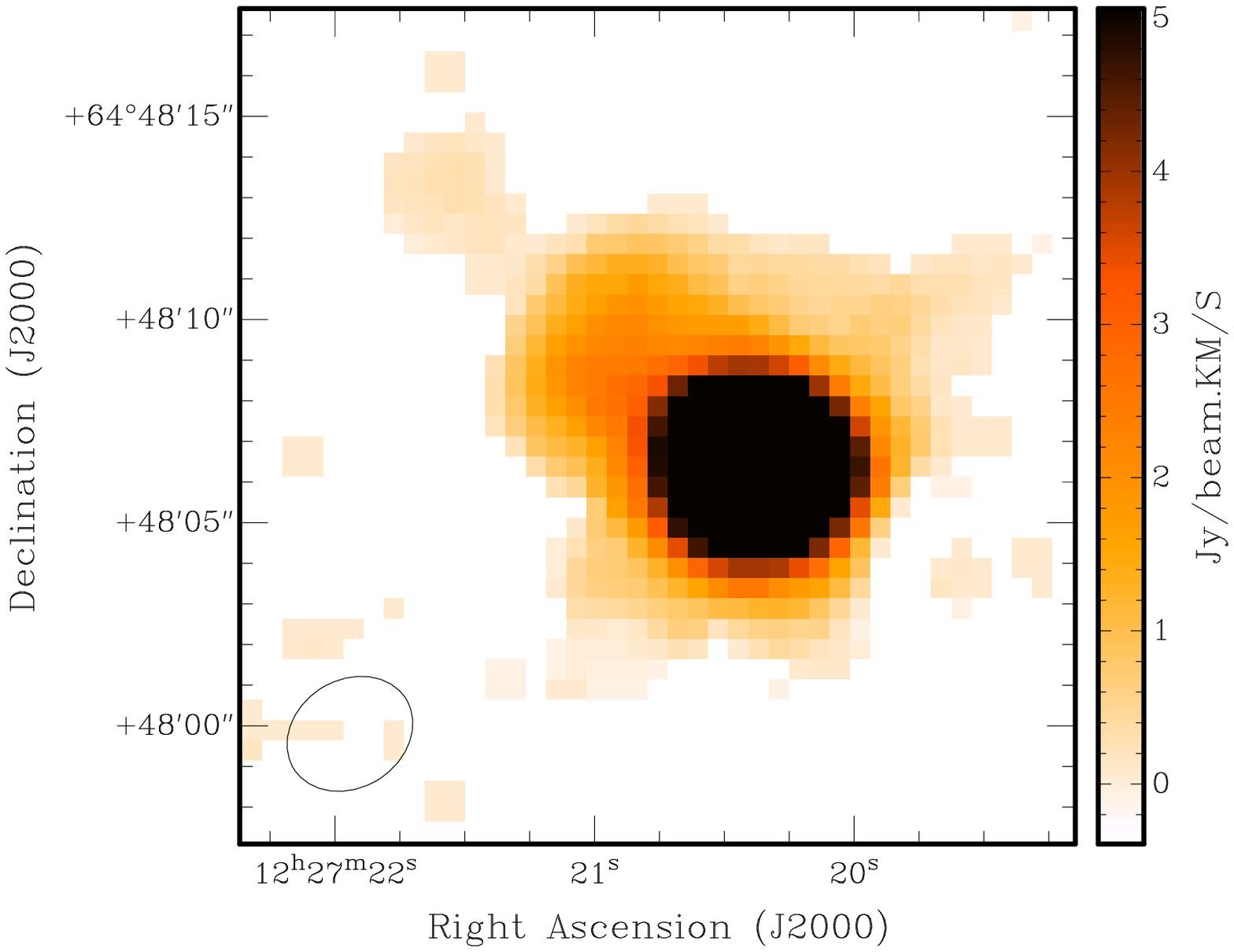}\\
\includegraphics[angle=0,width=8cm]{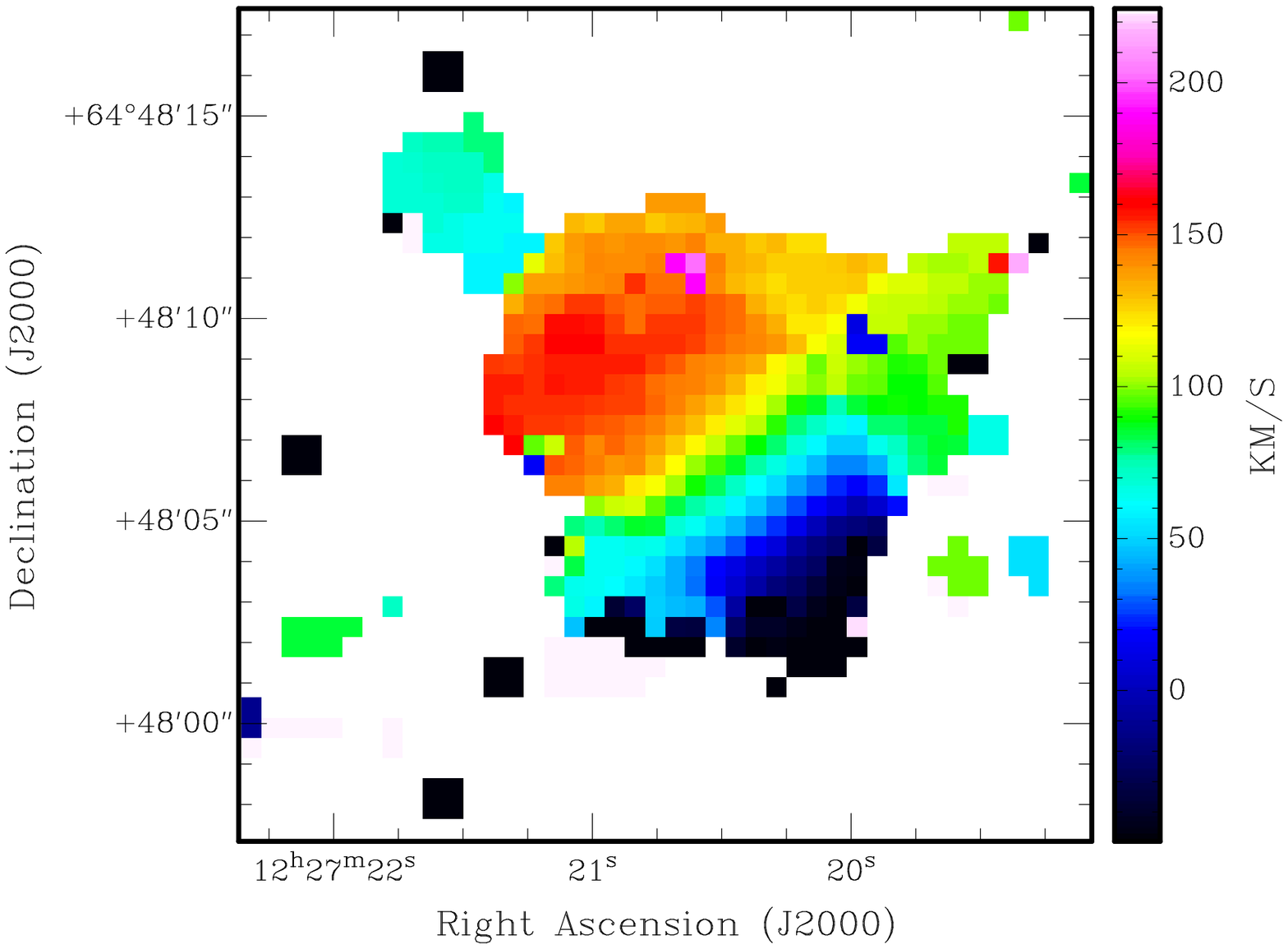}
      \caption{Plateau de Bure maps of the central CO disc in NGC\,4441: (top)
the integrated intensity map, (bottom) the velocity map. The beam is shown in 
the lower left corner of the upper panel. 
              }
         \label{pdb}
   \end{figure}

\begin{figure}[h!]
   \centering
   \includegraphics[angle=0,width=8cm]{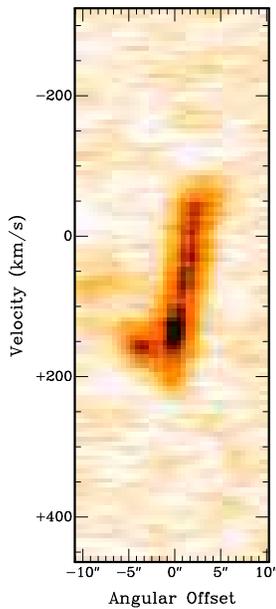}
      \caption{pv-diagram along the major axis of the rotating CO disc. 
The position angle is PA=222$\degr$, i.e. positive offsets are in south-western direction. The central velocity is 2652\,km\,s$^{-1}$.
              }
         \label{pv}
   \end{figure}

\begin{figure}[h!]
   \centering
   \includegraphics[angle=0,width=8cm]{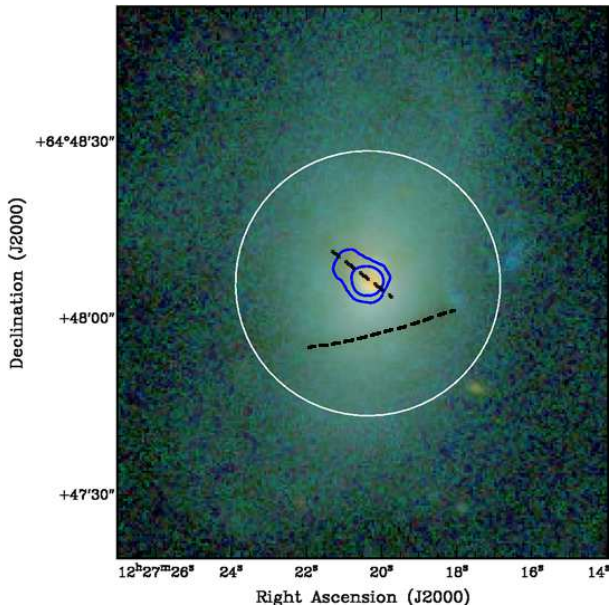}
      \caption{SDSS $irg$-colour composite image of the main body of NGC\,4441.
Various dust lanes are visible. Overlayed are two contours of the CO 
distribution to highlight the extended and the compact component. The two dashed lines mark two dust lanes. Note that the central dust lane is associated with the molecular gas. This dust lane
is broadend at the East side where a faint second peak in CO is seen. The 
white 
circle marks the primary beam of the Plateau de Bure interferometer.
              }
         \label{ima_optco}
   \end{figure}

In Fig.\,\ref{pdb} the integrated interferometer CO(1-0) intensity and 
velocity field maps of 
NGC\,4441
are shown. Since interferometers filter out emission from extended diffuse 
gas, the maps show the distribution of fairly dense CO. We found what 
appears to be a small, inclined,
nuclear disc aligned with the dust lane seen in the optical. The position angle of both the dust lane and the disc are identical. The orientation
of the disc is northeast--southwest. Thus, it is inclined compared with the optical
major axis. Most of the gas is concentrated around the optical nucleus. A
 second, fainter component with a small velocity dispersion is found north-east
of the centre seen in Figs.\,\ref{pdb},\ref{co_chanmap} ('hook' in Fig.\,\ref{pv}). This second peak
is aligned with a broadened dust feature seen in the optical image 
(Fig.\,\ref{ima_optco}).\\
It is worth to note here that besides the central dust lane associated with
the bulk of molecular gas, there are several faint off-centered dust lanes
visible in optical images (e.g., south of the centre, Fig.\,\ref{ima_optco}).
These lanes lie within the primary beam of  the PdBI. Thus, if there was
significant molecular gas also associated with these dust lanes, we would have 
detected it.\\
The total gas mass of the disc was calculated as $M_{H2} =
0.92\cdot10^4 \cdot F_{CO} \cdot D^2 = 4.1 \cdot 10^8\,{\rm M_{\odot}}$, with $D$ denoting the distance in Mpc, $F_{\rm }$ being the 
integrated CO flux in Jy\ km\ s$^{-1}$.  Thus, with the interferometric observations we detected $\sim 87\%$ of the molecular gas mass seen in single-dish observations.\\
If we interpret the central feature as an inclined disc, its position-velocity
diagram can be interpreted as the regular rotation pattern of a solid body
(Figs.\,\ref{pdb},\ref{pv}).
The main parameters of this feature are given in Table\,\ref{disc}. 
We calculated the dynamical mass of the nuclear disc with $v_{\rm rot,max} = 
100\,{\rm km\ s^{-1}}$ at a radius of $R = 5\arcsec = 875$\,kpc. The disc 
appears
to be close to edge-on, so we assume an inclination of $i = 80\degr$. An error
of $\pm 10\degr$ in inclination does not affect the result significantly. Thus,
we calculate the dynamical mass as $M_{\rm dyn} ({\rm M_{\odot}})= R\cdot (v_{\rm rot,max}
/sin\ i)^2\ G^{-1} = 2.1\cdot 10^9$\,\msun.

\begin{table}[h!]
\caption{Properties of the nuclear molecular gas feature found in NGC\,4441.} 
\label{disc}
\centering
\begin{tabular}{lc}
\hline\hline
property & value  \\
\hline
size (\arcsec /kpc)  & 10.3$\times$6.4/1.8$\times$1.1\\
PA ($^{\circ}$) & 222\\
$\Delta v$ ($\rm km\ s^{-1}$)& 295\\
$F_{CO}$ (Jy\,km\ s$^{-1}$) & 29.1 \\
$M_{\rm disc,H2}$ (\msun) & 4.1$\cdot 10^8$\\
$M_{\rm disc,dyn}$ (\msun) & 2.1 $\cdot 10^9$\\
\hline
\end{tabular}
\end{table}

\subsection{$\rm ^{13}$CO(1-0) and $\rm ^{13}$CO(2-1)}
\begin{figure}[h!]
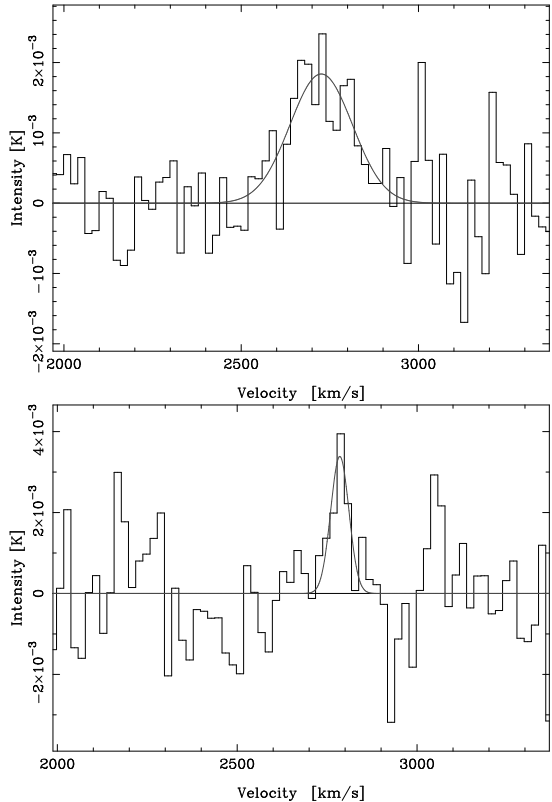

   \centering
   \includegraphics[angle=270,width=7.2cm]{13co1-0Tmb.ps}
    \includegraphics[angle=270,width=7.2cm]{13co2-1Tmb.ps}
      \caption[NGC\,4441: $\rm ^{13}$CO spectra]{$\rm ^{13}$CO observed at the
              center position of 
              NGC\,4441, 
              measured
              with the IRAM 30\,m  telescope. The intensity is given in units of $
              T_{\rm MB}$. {\bf top:} J=1--0, rms=0.7\,mK. {\bf bottom:} 
              J=2--1, rms=1.3\,mK. Overlaid are Gaussian fits to the lines. 
The parameters 
              are given in Table\,\ref{n4441chap_13cotab}.
              }
         \label{n4441chap_13co}
   \end{figure}

We also observed NGC\,4441 in $\rm ^{13}$CO(1-0) and
$\rm ^{13}$CO(2-1) at the center position. The $\rm ^{13}$CO(1-0) line is
clearly detected, the 
$\rm ^{13}$CO(2-1) is only tentatively detected at a 3\,$\sigma$ level, but
 a feature at the expected velocity 
of 2790\,km\ s$^{-1}$ is found (see
Fig.\,\ref{n4441chap_13co}). Table\,\ref{n4441chap_13cotab} gives the 
linewidths and integrated intensities 
of all four lines at the center position.  We
fitted only one single Gaussian to the $\rm ^{13}$CO lines. The fitted center
velocity of  $\rm ^{13}$CO(1-0)
(Table\,\ref{n4441chap_13cotab}) lies between the derived
values for the 
blue- and redshifted line using the $\rm ^{12}$CO transitions.
In contrast, in $\rm ^{13}$CO(2-1) we
see (if at all) the component with the higher velocity, which is also the
stronger one in $\rm ^{12}$CO at the center. 
However, the signal-to-noise ratio of $\rm ^{13}$CO(2-1) is too low to
draw firm conclusions.
\\
In Table\,\ref{n4441chap_coprop} the derived molecular gas mass and
the molecular line intensity ratios based on the integrated
intensities given in Table\,\ref{n4441chap_13cotab} are listed.

\begin{table*}[h!]
\caption[NGC\,4441: $\rm ^{12}$CO and $\rm ^{13}$CO at the center]{Parameters of $\rm ^{12}$CO and $\rm ^{13}$CO at the central
  position. Gaussians were 
  fitted to determine linewidths, peak intensity and center velocity. For
  $\rm ^{12}$CO, we 
  fitted two components to the measured line, in $\rm ^{13}$CO only one
  component was 
  fitted. The given intensity is integrated over both
  components. Temperatures are given in $ T_{\rm MB}$.}            
\label{n4441chap_13cotab}      
\centering                       
\begin{tabular}{lccccccccc}       
\hline\hline            
line & \multicolumn{2}{c}{center velocity (lsr)} &
\multicolumn{2}{c}{linewidth}  & \multicolumn{2}{c}{Peak} &
int. Intensity \\ 
& \multicolumn{2}{c}{(km\ s$^{-1}$)} & \multicolumn{2}{c}{(km\ s$^{-1}$)} &
\multicolumn{2}{c}{(K)}  & (K\,km\ s$^{-1}$)\\
  \hline                  
$\rm ^{12}$CO(1-0)&  2698 & 2796 &125 & 147 & 0.02 & 0.036& 7.52 $\pm$
0.22\\
$\rm ^{12}$CO(2-1)& 2655 & 2802&93 & 114& 0.016, & 0.039& 6.10 $\pm$ 0.73\\
$\rm ^{13}$CO(1-0)& \multicolumn{2}{c}{2725}&
  \multicolumn{2}{c}{208} & \multicolumn{2}{c}{0.0018}&
      0.37 $\pm$ 0.06\\ 
$\rm ^{13}$CO(2-1)&  \multicolumn{2}{c}{2785}&\multicolumn{2}{c}{61}&
\multicolumn{2}{c}{0.0033} & 0.25 $\pm$ 0.09\\
\hline                                   
\end{tabular}
\end{table*}

\begin{table}[h!]
\caption[NGC\,4441: Molecular line ratios]{Integrated intensity line ratios derived
  from the CO measurements at the central position. In brackets the
errors of the line ratio are given, following standard error propagation
calculations. A source size of 22\arcsec is used for the line ratios.} 
\label{n4441chap_coprop}
\centering
\begin{tabular}{lc}
\hline\hline
 property& result\\
\hline
 $M_{\rm H2}$ ($10^8$ \msun) & 4.6\\
$^{12}$CO(2-1)/$^{12}$CO(1-0)& 0.5 (0.06)\\
$^{13}$CO(2-1)/$^{13}$CO(1-0) & 0.42 (0.17)\\
$^{12}$CO(1-0)/$^{13}$CO(1-0) & 21 (3.5)\\
$^{12}$CO(2-1)/$^{13}$CO(2-1) & 25 (7.6)\\
\hline
\end{tabular}
\end{table}

\section{Radiative transfer calculations}
\label{radtrans}
To investigate the  physical conditions of the ISM, we used
{\tt RADEX}\footnote{http://www.strw.leidenuniv.nl/$\sim$moldata/radex.html} 
which is a
  one-dimensional spherically symmetric non-LTE radiative transfer
code available on-line 
  \citep{radexman,2005A&A...432..369S,2007A&A...468..627V}. In 
this code, the mean escape
  probability method 
  for an isothermal and homogeneous medium is used for the calculations.  

\begin{figure*}[h!]
   \centering
   \includegraphics[angle=-90,width=8cm]{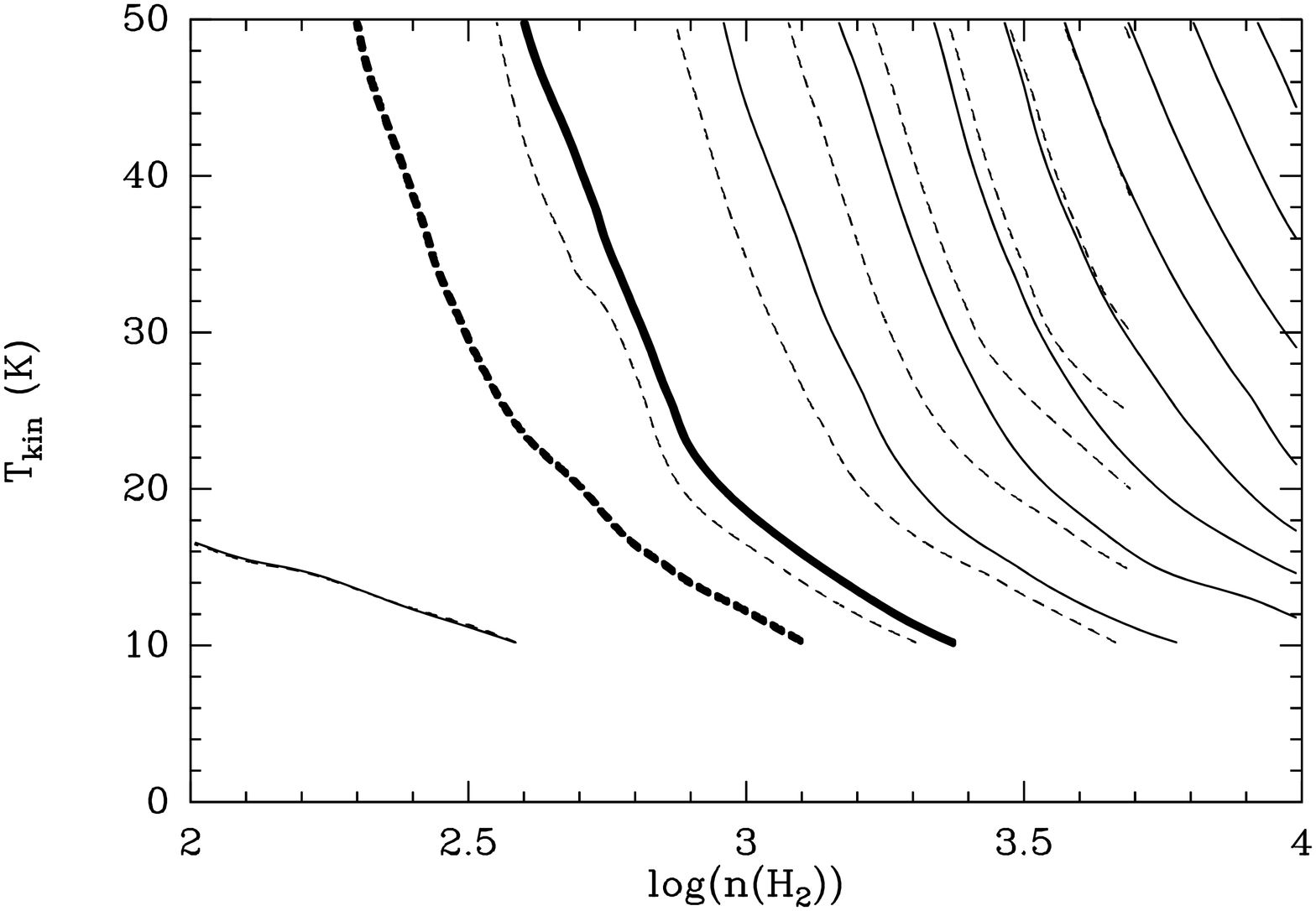}
\includegraphics[angle=-90,width=8cm]{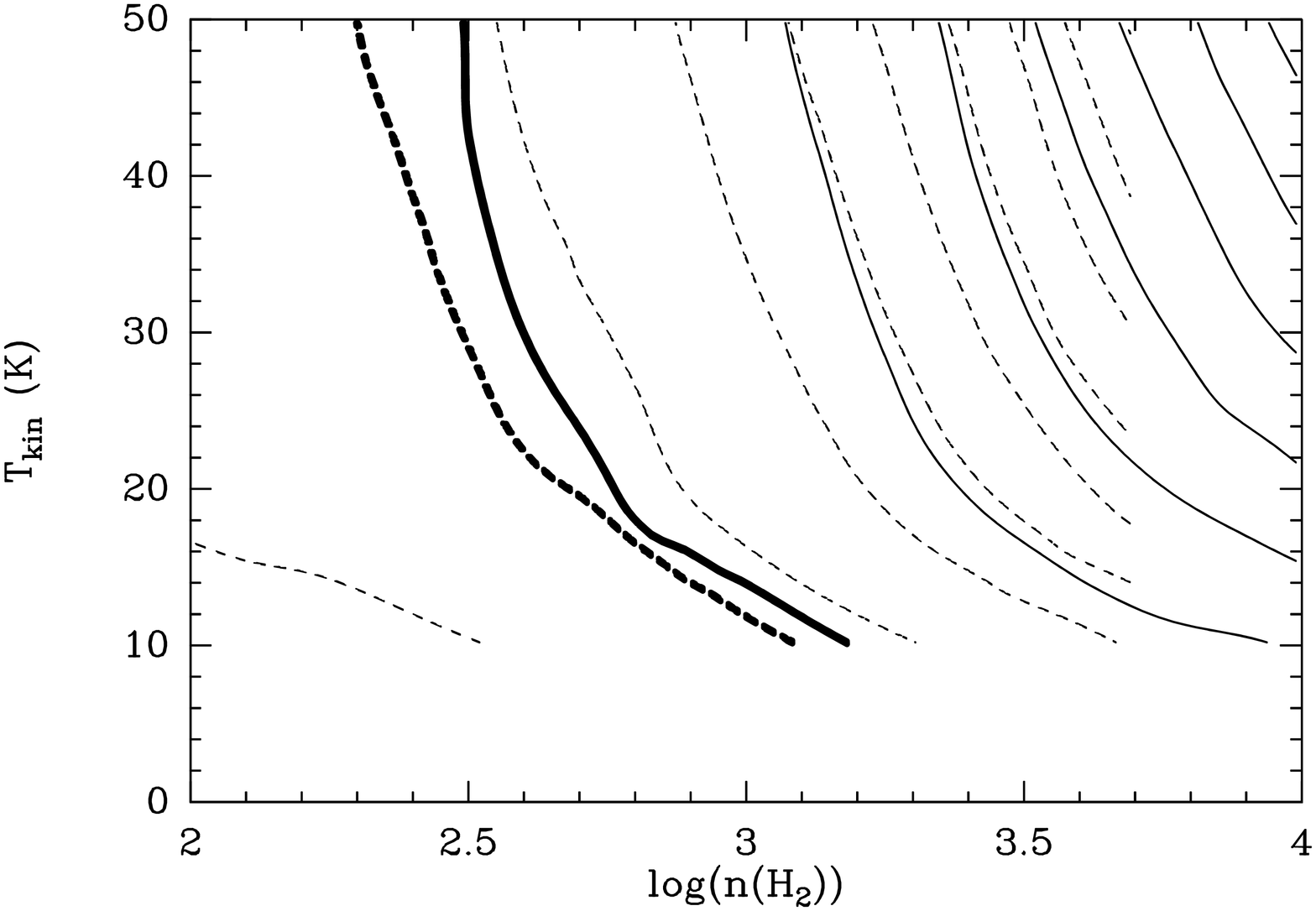}
 \caption[NGC\,4441: RADEX results]{Results of the {\tt RADEX} 
             analysis. The $\rm
            H_2$ column density is $\rm 10^{20}\,cm^{-2}$ (right), and
            $\rm 10^{21}\,cm^{-2}$. Shown are the line ratios of
            $^{12}$CO(2--1)/$^{12}$CO(1--0) 
            (solid) and $^{13}$CO(2--1)/$^{13}$CO(1--0) (dashed). The bold lines
            mark the measured line ratios of NGC\,4441. The levels of the line
            ratios start
            at 0.25 and proceed in steps of 0.25. For this analysis,
we assumed a linewidth of 10\,km\ s$^{-1}$ per cloud and abundances of $\rm
X[^{12}CO]=10^{-4}$ and $\rm X[^{13}CO]=2\cdot10^{-6}$
            }
         \label{n4441chap_radexlvgresult}
   \end{figure*}

We assume that the $\rm ^{12}$CO lines are optically thick ($\rm \tau > 1$), so
that 
the CO is selfshielded from the UV radiation field. 
The high $^{12}$CO/$^{13}$CO line ratio limits the optical
depths to  
$\tau < 10$, unless the $^{13}$CO abundances are extraordinary low. 
Furthermore, we assume 
"standard" abundances of $^{12}$CO and $^{13}$CO of $\rm
X[^{12}CO]=10^{-4}$ and $\rm 
X[^{13}CO]=2\cdot 10^{-6}$ \citep[e.g.,][]{1987ApJ...315..621B,
2000A&A...358..433M}.
The line ratios fitted by the radiative transfer model constrain the
average properties of an ensemble of clouds that are not filling the beam.
We are assuming that the emission is emerging from a single cloud type - which
in reality is very unlikely to be true, but provides us with a first 
handle on the
main properties of the $^{12}$CO/$^{13}$CO emitting gas.
\\
We explore a range of $\rm H_2$ column densities and temperatures of
$\rm 10^{18}-10^{22}\,cm^{-2}$ and
10--50\,K. 
In Fig.\,\ref{n4441chap_radexlvgresult} the results  for
$\rm H_2$ column
densities of $\rm 10^{20}\,cm^{-2}$ and $\rm 10^{21}\,cm^{-2}$ are displayed.
The levels decrease
from high temperatures at low densities to low temperatures at high
densities. The measured line ratios $^{12}$CO(2-1)/$^{12}$CO(1-0) and
$^{13}$CO(2-1)/$^{13}$CO(1-0) both indicate subthermally excited
emission, i.e., 
either low temperatures and/or low density.\\
The best
agreement of the $^{12}$CO(2--1)/$^{12}$CO(1--0) and
$^{13}$CO(2--1)/$^{13}$CO(1--0) 
lines ratios is found for a column density per cloud of $\rm 10^{21}\,cm^{-2}$.\\
  From the $^{12}$CO and $^{13}$CO ratios alone it is not
 possible 
 to determine the temperature in a more specific way, but as seen in
 Fig.\,\ref{n4441chap_radexlvgresult}, for higher temperatures ($\rm > 20\,K$)
 the 
 density is well below 1000\,$\rm cm^{-3}$. 
There are no signs of warm, dense ($ n>10^4\,$cm$^{-3}$) cores. For
the gas to be dense 
and still fit the observed line ratios, temperatures of 5\,K are
required, which is 
significantly lower than the dust temperature. This is of course
technically possible, but 
such a cold dense component would quickly collapse and form stars. We
find it more 
likely 
 that the observed line ratios indicate a low
density ($n < 1000$ cm$^{-3}$) 
molecular ISM where the clouds are diffuse --- i.e. not entirely
self-gravitating --- or consisting 
of extended outskirts of much smaller, bound clouds. This would be in agreement with the moderate to low  star formation
rate derived for NGC\,4441 \citep{n4441hi}. \\
Interestingly, the $^{12}$CO/$^{13}$CO line ratio in NGC\,4441 is unusually high for
a galaxy with a moderate to low star formation rate. Values found
here are more reminiscent of starburst nuclei (15--20) \citep{1995A&A...300..369A}. \\
It is possible that the ISM in NGC\,4441 is the left-over from a burst of star
formation that consumed the denser fraction of the gas. Diffuse gas may also
be the result of dynamical impact on the gas properties: tidal forces,
bar-induced shocks or the merger event.
\section{Discussion}

\subsection{A counter-rotating nuclear molecular disc?}
Nuclear rotating discs are often found in interacting galaxies 
\citep[e.g.,][]{1996ApJ...457..678B,1998ApJ...507..615D,
1999AJ....117.2632B,2001ApJ...550..104Y}. Even though these studies 
concentrate on ULIRGs, some of the observed features are similar to 
the disc found in NGC\,4441. The discs
found in mergers of spirals are very compact, with typical radii of 300--800\,pc.
 Furthermore, the molecular gas fraction traced by CO  tends to have a  rather low  density ($\rm 10^2 - 
10^3\,cm^{-3}$) \citep{1998ApJ...507..615D}. 
Therefore, the CO lines are subthermally excited \citep{1998ApJ...507..615D}, as
 we also assume for NGC\,4441. However, in ULIRGs 
turbulence is rather high (up to 100\, $\rm km\ s^{-1}$, \citep[e.g.,][]{1998ApJ...507..615D}) which explains 
why gravitational instabilities occur leading to the observed high star 
formation rate. The ISM in the nuclear disc of NGC\,4441 might be less 
turbulent due to a different merger history. If NGC\,4441 is a S+E merger, 
only one partner brings in a significant amount of molecular gas, which might 
form an interaction-triggered disc more smoothly. Therefore, the disc might be 
more stable and thus only moderate ongoing star formation is observed. \\
Besides the diffuse component generally found in ULIRGs, the bulk of 
the molecular gas is in a dense phase ($\rm 10^5 - 10^6\,cm^{-3}$), 
supporting the extreme star formation \citep[e.g.,][]{1992ApJ...387L..55S,2006astro.ph.10378G}. Based on our radiative transfer modelling we find no hints of such a dominant dense gas phase in NGC\,4441, in accordance to the moderate star formation rate. \\
Compared with the rotation of the large-scale HI disc \citep{n4441hi}, the 
nuclear molecular disc is kinematically decoupled. Its rotation axis is shifted 
by $\sim 100^{\circ}$. \\
Kinematically decoupled cores (KDCs) are found in various galaxies, in spirals
as well as in S0 and ellipticals \citep[e.g.,][]{2008MNRAS.390...93K,
2000AJ....120..703H,1996MNRAS.283..543K}.
 First simulations of \cite{1991Natur.354..210H} showed that 
counter-rotating nuclear discs can occur in an equal-mass disc-disc merger
in which the two progenitors have anti-parallel spins (i.e., a {\it retrograde}
merger). In this scenario, the nuclear disc contains gas from the
centres of the original gas discs which was exposed to strong gravitational
torques and thus losing most of the angular momentum. Because of that, no
information about the original sense of rotation is preserved and thus the 
rotation axis of the new disc is not related to the original ones. In contrast,
the gas in the outer parts can retain its angular momentum and thus conserves
the original motion. In this scenario, the merger remnants evolve into elliptical
galaxies which are often found to host a KDC.\\
Further theoretical studies showed that KDCs in S0 and spirals can occur due
to minor mergers with gas rich companions, and even mergers with at least one
elliptical can lead to (stellar) KDCs \citep[e.g.,][]{1999IAUS..186..149B,
1998ApJ...506...93T,2000AJ....120..703H,2008A&A...477..437D}.
 Besides the variety of merging partners leading to a 
KDC, all simulations have in common that the merger geometry has to be 
retrograde. 

\subsubsection{Is the minor axis dust lane
really an inclined disc{\bf ?}}
 The interpretation of the crossing central dust lane of NGC\,4441 as an
inclined disc seems likely due to the large velocity shift along
the lane as well as the regular-looking shape of the associated {\it pv}
diagram. The disc appears stable against star formation and may be
filled with diffuse, unbound molecular gas. However, one problem
is the apparent counter-rotation of the inclined disc. In our
previous paper on \hi\ \citep{n4441hi} we suggested that the interaction
 between the
two galaxies was pro-grade, because of a large amount of angular 
momentum still remaining in the gas of the tidal tails. However, retrograde
mergers appear to
be a requirement to produce counter-rotating cores. If the pro-grade
scenario were correct, there
is a problem with the interpretation of the dust lane as an inclined disc.
A final answer cannot be given by these
observations alone but is only possible
by obtaining numerical simulations.\\
Minor axis dust lanes in other E+S mergers (such as Cen-A, Fornax-A,
NGC\,4194 (the Medusa)) present a variety of dynamical behaviour. In
the Medusa the dust lane is morphologically similar to the one in NGC\,4441,
but has a much smaller velocity gradient and is not interpreted as a disc,
but rather as a lane along which gas is fed to the nucleus \citep{2000A&A...362...42A}. In 
NGC\,4441, however, we do not find such a central starburst.

 \subsection{Present state and expected future evolution of NGC\,4441}
The presence of an extended molecular gas reservoir in NGC\,4441 is comparable
to what is found in NGC\,4194, the prototypical S+E merger candidate 
\citep{2000A&A...362...42A}. 
However, the geometry compared to the optical appearance is 
different. In the Medusa the CO follows roughly the direction
of the optical tidal tail, whereas in NGC\,4441 the molecular gas seems to be
 more
extended to 
the opposite side of the optical tail. This is probably a result of a different merger geometry of these two galaxies.
In any case, extended molecular gas indicates a different merger history compared
to ULIRGs in which molecular gas is concentrated in the central kpc. Based on
the similar optical morphology, in particular the strength of the features,
it seems likely that NGC\,4441 and NGC\,4194 are in a similar merger state.\\
 Interestingly, the molecular gas mass of NGC\,4441 is only a quarter
of the Medusa ($\rm 4.6 \cdot 10^8 \,M_{\odot}$ for NGC\,4441), 
whereas the amount of atomic hydrogen
is comparable  
($\rm
1.46 \cdot 10^9\,M_{\odot}$ in NGC\,4441 versus $\rm 2 \cdot
10^9\,M_{\odot}$ in the 
Medusa) \citep{n4441hi,medusa}. The question arises, why the starburst 
in NGC\,4441 has already
faded, while the Medusa is still intensely forming
stars. 
The relation between the ongoing star formation  and the
available resource for 
star formation (i.e., dense gas) can be expressed by the star formation
efficiency: SFE($\rm 
  \frac{1}{yr}$) = $\rm 
  \frac{SFR_{FIR}}{M_{H2}}$.
Thus, for NGC\,4441 we derive a star formation efficiency of $\rm SFE =
2.1\cdot 10^{-9}\,yr^{-1}$. The equivalent gas depletion time is $\rm \tau =
1/SFE = 4.8\cdot 10^8\,yr$.   
If we assume a conversion factor $\rm X_{CO}$ to be similar as in
the Medusa, we can directly compare the SFE with that in NGC\,4194,
because their metalicities are similar \citep{optsample,1996PASJ...48..275A}.
Comparing NGC\,4441 with the very effective star burst in the Medusa (up to an
efficiency of $\rm 1.7\cdot 10^{-8}\,yr$ \citep{2000A&A...362...42A}),
NGC\,4441 has 
  a moderate star formation efficiency:  $\rm SFE\,(NGC\,4194) \sim 10\times
  SFE\,(NGC\,4441)$. 
Since the $\rm H_2$-to-CO conversion factor might be different for starbursting and 'quiescent' galaxies, we may
overestimate the mass, and therefore underestimate the SFE in comparison with the Medusa. But these dependencies are not well understood so far and can't be taken into account here.
From the radiative transfer models, we find that the
  molecular gas is rather thin and cold in NGC\,4441, which 
  makes it difficult to form new stars, whereas in the Medusa 
the CO lines are thermalised, indicating higher average gas densities and/or
temperatures. It is, however, unclear, whether the
  differences are due to ageing effects, e.g., NGC\,4441 might
  have transformed all dense gas into stars, whereas in the Medusa this
  process is still ongoing. Furthermore, differences in the gas
  densities can be explained by different gas reservoirs provided by the progenitors
  and/or different merger geometry.   \\
Thus, if this galaxy is already in a
post-starburst phase, it still manages to retain a large amount of gas
in principle available for 
star formation. \\
This is in agreement with observations of so-called
'E+A' galaxies, i.e. post-starburst with an optical spectrum
reflecting a strong A star population
superimposed on an old elliptical-like population.
A small but significant fraction of E+A galaxies have optical tidal
features, suggesting a 
merger event only 0.5--1\,Gyr ago \citep[e.g.,][]{1996ApJ...466..104Z,
2004MNRAS.355..713B}. Furthermore, \cite{2006ApJ...649..163B}
obtained HI 
observations of a sample of E+A 
galaxies and show that the total atomic gas content may not be consumed
until a galaxy reaches that post-starburst phase. \\
Furthermore, a large amount of diffuse molecular gas can survive a
starburst phase \citep{2002PASJ...54..541K}, when the gas in the
dense cores is consumed during the star formation process. It is
argued by \cite{2002PASJ...54..541K} that the remaining molecular gas
can be stable against gravitational instabilities. Such instabilities could
lead to 
a condensation and thus trigger star formation. In particular in
early-type galaxies such as NGC\,4441, the mass density increases steeply towards
the center, which leads to a rise of the rotation velocity. Therefore,
the epicyclic frequency, which is proportional to the critical mass
density, also rises and hence also the critical mass density, which
is the lower threshold density for possible star formation. In other
words, the star formation is suppressed, although raw material is
still present in the galaxy. \\
Indeed, the small regularly rotating central molecular gas feature which
is likely a disc, seems to be stable, i.e., not star forming and
thus is similar to what is found in NGC\,5195 by \cite{2002PASJ...54..541K}.


\section{Summary}

\begin{enumerate}
\item We observed the advanced merger remnant \,NGC4441 in $^{12}$CO(1--0) 
      using the
      Onsala 20\,m telescope. We determined a 
      total molecular gas mass of ${\rm 4.6 \cdot 10^8\,M_{\odot}}$.
\item The star formation efficiency is $\rm 2.1\cdot 10^{-9}\,yr^{-1}$
      and the gas depletion time $\tau = 4.8 \cdot10^8$\,yr.
\item Following up, we mapped NGC\,4441 with the single-dish IRAM 30\,m
      telescope in $^{12}$CO(1--0) and $^{12}$CO(2--1). We found extended 
      molecular gas out
      to 22\arcsec (3.9\,kpc). 
\item Two distinct velocity components with a velocity difference of $\rm 
      \sim 100\,km\ s^{-1}$ were detected. The relative intensities of both 
      components
      vary with distance from the center and with transition.
\item High resolution imaging using Plateau de Bure revealed a small central
      rotating molecular gas disc hosting most of the molecular gas in 
NGC\,4441. In contrast to ULIRGs, which also have compact molecular gas 
disc fueling the central superstarburst, the disc in NGC\,4441 seems to be
 stable and thus does not support star formation. 
\item The central molecular gas disc has a different sense of rotation than the large-scale \hi distribution. This is a strong indicator of a kinematically decoupled core.
\item Furthermore, we observed the $^{13}$CO(1--0) and $^{13}$CO(2--1) line
      with the IRAM 30\,m telescope to estimate the molecular gas properties 
      using the
      radiative transfer model {\tt RADEX}.
      We derived $^{12}$CO(2--1)/$^{12}$CO(1--0) and
      $^{13}$CO(2--1)/$^{13}$CO(1--0) line ratios which are consistent with a
      diffuse ($ n_{\rm H2} \leq 1000\,{\rm cm^{-3}}$) molecular medium. However,
      the
      $^{12}$CO(1--0)/$^{13}$CO(1--0) ratio is unusually high
      ($^{12}$CO(1--0)/$^{13}$CO(1--0) = 21), which is typical for the inner
      centers of luminous starbursts \citep{1995A&A...300..369A}. 
\item The moderate star formation rate of 1--2 $\rm M_{\odot}\ yr^{-1}$ 
      is in good agreement with the
      results from the molecular gas analysis. Because not much dense 
      ($ n \geq 10^4$\,cm$^{-3}$)  gas is 
      present,
      no enhanced ongoing star formation can happen. 
       The strength of a past starburst remains unclear, however.
\item NGC\,4441 may be a local candidate for an E+A galaxy, still hosting a significant amount of gas but the starbust has been faded away already. 
\end{enumerate}

\begin{acknowledgements}
 We
thank Evert Olsson for his assistance of the IRAM observations. We thank the
Onsala Space Observatory, and IRAM staff for their support during the
observations. 
  This research has made use of the NASA/IPAC Extragalactic
 Database (NED) which is operated by the Jet Propulsion Laboratory, California
 Institute of Technology, under contract with the National Aeronautics and
 Space Administration. The research was partially supported by the German
 Science 
 Organisation (DFG) through the Graduiertenkolleg 787.

\end{acknowledgements}

\bibliographystyle{aa}
\bibliography{manthey}

\end{document}